\PassOptionsToPackage{unicode}{hyperref}
\PassOptionsToPackage{hyphens}{url}
\PassOptionsToPackage{dvipsnames,svgnames,x11names}{xcolor}
\documentclass[
  11pt,
  a4paper,
]{article}
\usepackage{xcolor}
\usepackage[margin=1in]{geometry}
\usepackage{amsmath,amssymb}
\usepackage{iftex}
\ifPDFTeX
  \usepackage[T1]{fontenc}
  \usepackage[utf8]{inputenc}
  \usepackage{textcomp} 
\else 
  \usepackage{unicode-math} 
  \defaultfontfeatures{Scale=MatchLowercase}
  \defaultfontfeatures[\rmfamily]{Ligatures=TeX,Scale=1}
\fi
\usepackage[]{mathpazo}
\ifPDFTeX\else
\fi
\IfFileExists{upquote.sty}{\usepackage{upquote}}{}
\IfFileExists{microtype.sty}{
  \usepackage[]{microtype}
  \UseMicrotypeSet[protrusion]{basicmath} 
}{}
\usepackage{setspace}
\makeatletter
\@ifundefined{KOMAClassName}{
  \IfFileExists{parskip.sty}{%
    \usepackage{parskip}
  }{
    \setlength{\parindent}{0pt}
    \setlength{\parskip}{6pt plus 2pt minus 1pt}}
}{
  \KOMAoptions{parskip=half}}
\makeatother
\usepackage{longtable,booktabs,array}
\usepackage{calc} 
\usepackage{etoolbox}
\makeatletter
\patchcmd\longtable{\par}{\if@noskipsec\mbox{}\fi\par}{}{}
\makeatother
\IfFileExists{footnotehyper.sty}{\usepackage{footnotehyper}}{\usepackage{footnote}}
\makesavenoteenv{longtable}
\usepackage{graphicx}
\makeatletter
\newsavebox\pandoc@box
\newcommand*\pandocbounded[1]{
  \sbox\pandoc@box{#1}%
  \Gscale@div\@tempa{\textheight}{\dimexpr\ht\pandoc@box+\dp\pandoc@box\relax}%
  \Gscale@div\@tempb{\linewidth}{\wd\pandoc@box}%
  \ifdim\@tempb\p@<\@tempa\p@\let\@tempa\@tempb\fi
  \ifdim\@tempa\p@<\p@\scalebox{\@tempa}{\usebox\pandoc@box}%
  \else\usebox{\pandoc@box}%
  \fi%
}
\def\fps@figure{htbp}
\makeatother
\NewDocumentCommand\citeproctext{}{}
\NewDocumentCommand\citeproc{mm}{%
  \begingroup\def\citeproctext{#2}\cite{#1}\endgroup}
\makeatletter
 \let\@cite@ofmt\@firstofone
 \def\@biblabel#1{}
 \def\@cite#1#2{{#1\if@tempswa , #2\fi}}
\makeatother
\newlength{\cslhangindent}
\newlength{\csllabelwidth}
\newenvironment{CSLReferences}[2] 
 {\begin{list}{}{%
  \setlength{\itemindent}{0pt}
  \setlength{\leftmargin}{0pt}
  \setlength{\parsep}{0pt}
  \ifodd #1
   \setlength{\leftmargin}{\cslhangindent}
   \setlength{\itemindent}{-1\cslhangindent}
  \fi
  \setlength{\itemsep}{#2\baselineskip}}}
 {\end{list}}
\usepackage{calc}

\usepackage{lineno} 
\usepackage{amsmath} 
\usepackage{float} 

\usepackage{lscape}
\newcommand{\blandscape}{\begin{landscape}}
\newcommand{\elandscape}{\end{landscape}}

\newcommand{\bleft}{\begin{flushleft}}
\newcommand{\eleft}{\end{flushleft}}

\usepackage{tikz}
\usetikzlibrary{positioning}
\usetikzlibrary{calc}

\usepackage[normalem]{ulem}

\usepackage{bookmark}
\IfFileExists{xurl.sty}{\usepackage{xurl}}{} 
\hypersetup{
  pdftitle={Bayesian decision theory for wildlife management under uncertainty: from inference to action},
  pdfauthor={Olivier Gimenez1*; Abigail G. Keller2; Cyril Milleret1},
  colorlinks=true,
  linkcolor={RoyalBlue},
  filecolor={Maroon},
  citecolor={Blue},
  urlcolor={RoyalBlue},
  pdfcreator={LaTeX via pandoc}}

\title{Bayesian decision theory for wildlife management under uncertainty: from inference to action}
\author{Olivier Gimenez\textsuperscript{1}* \and Abigail G. Keller\textsuperscript{2} \and Cyril Milleret\textsuperscript{1}}
\date{2026-08-25}

\begin{document}
\maketitle

\setstretch{2}
\small

\textsuperscript{1} CEFE, Univ Montpellier, CNRS, EPHE, IRD, Montpellier, France
\textsuperscript{2} University of California, Berkeley, Berkeley, California, USA

\texttt{*} Corresponding author: \href{mailto:olivier.gimenez@cefe.cnrs.fr}{\nolinkurl{olivier.gimenez@cefe.cnrs.fr}}

\normalsize

\paragraph{Acknowledgements:}

This research was supported by the ExposUM Institute of the University of Montpellier and the ANR through the project NACHOS for ``Interdisciplinary approach to small carnivores - humans relationships'' (grant ANR-25-CE03-5469).

\paragraph{Declaration of Competing Interest:}

The author declares that he has no known competing financial interests or personal relationships that could have appeared to influence the work reported in this paper.

\paragraph{Declaration of generative AI and AI-assisted technologies in the writing process:}

During the preparation of this work, the author used ChatGPT to polish the text and enhance the English language. After using this tool, the author reviewed and edited the content as needed and take full responsibility for the content of the published article.

\paragraph{Data availability statement:}

Codes and data are available at \url{https://github.com/oliviergimenez/bayesian-decision-theory/}.

\paragraph{Author contributions:}

O. Gimenez conceived the study, developed the methodology, conducted the analyses, and wrote the original manuscript draft. C. Milleret and A.G. Keller contributed to the interpretation of the work and critically revised the manuscript. All authors read and approved the final version of the manuscript.

\newpage

\section*{Abstract}
\begin{enumerate}
\item Ecologists are increasingly expected to inform management decisions under uncertainty, yet most analytical workflows stop at statistical inference. This disconnect limits the practical impact of ecological modelling, particularly in high-stakes contexts such as wildlife management, where decisions must balance ecological, economic and social objectives.

\item Bayesian decision theory provides a coherent framework to bridge this gap. It propagates uncertainty from posterior distributions to quantify the consequences of alternative actions through utility functions. Despite its strong theoretical foundations, it remains underused in ecology.

\item Here, we present a practical workflow for implementing Bayesian decision theory using standard Bayesian tools. We illustrate the approach with two case studies. First, wolf management in France, where the decision consists of selecting the number of wolves that can be removed under uncertainty about population dynamics. Second, invasive muskrat management in the Netherlands, where the decision involves allocating a fixed control effort across space. In both cases, expected utility is computed by integrating over posterior uncertainty, explicitly accounting for management trade-offs.

\item Results show that optimal decisions emerge as a compromise between competing objectives. In the wolf case, optimal harvest balances removal benefits and population risk. In the muskrat case, optimal effort increases with the importance of population reduction and is unevenly allocated across provinces.

\item These examples show that Bayesian decision theory can be implemented as a direct extension of standard inference. It provides a formal interface between scientists, who characterize how the system works and quantify uncertainty, and decision-makers, who define objectives, values and trade-offs. By making these trade-offs explicit, it enhances transparency, reproducibility, and relevance for management. Bayesian decision theory also provides a common framework for reconnecting Bayesian statistics, decision analysis and risk analysis. These fields share theoretical foundations but have developed separately, and bringing them together could strengthen the link between ecological inference and action.
\end{enumerate}

\emph{Keywords}: Bayesian inference; Decision-making; Hierarchical models; Invasive species management; Uncertainty propagation.

\section{Introduction}\label{introduction}

Ecological systems are increasingly shaped by rapid environmental change, multiple interacting stressors and strong societal expectations. Ecologists are therefore increasingly expected not only to understand ecological processes, but also to inform management decisions under uncertainty. Such decisions are typically made with incomplete knowledge of system states and dynamics (\citeproc{ref-polaskyDecisionmakingGreatUncertainty2011}{Polasky et al. 2011}).

Over the past decades, Bayesian approaches have become central to ecological modelling, offering a coherent framework to quantify and propagate uncertainty through posterior distributions of model parameters and latent states (\citeproc{ref-hobbs2025}{Hobbs and Hooten 2025}). This shift has improved our ability to infer population abundance and trends, predict community dynamics, and account for imperfect detection (\citeproc{ref-RD2008}{Royle and Dorazio 2008}, \citeproc{ref-KR2020}{Kéry and Royle 2020}). However, in many applications, the analytical workflow effectively stops at inference: ecologists derive posterior distributions, assess uncertainty and compare models, but rarely proceed to the next step-formal decision-making. As a result, decision-makers are often left without quantitative guidance for choosing among actions.

Any formal decision analysis combines two fundamentally different components: knowledge about how the system works and judgments about what outcomes are desirable. Scientists and statisticians contribute to the former by estimating system states, processes and their uncertainty, whereas objectives, trade-offs and acceptable outcomes reflect the values of the decision-makers. Eliciting these values is a challenging part of decision analysis, and structured decision making provides well-developed tools for framing decisions, defining objectives and alternatives, and eliciting preferences (\citeproc{ref-Williams2002}{Williams et al. 2002}, \citeproc{ref-StructuredDecisionMaking2020}{Runge et al. 2020}).

Bayesian decision theory (\citeproc{ref-berger1985statistical}{Berger 1985}) provides a principled and conceptually simple framework to combine uncertainty --- through posterior distributions --- with the consequences of alternative actions --- through utility functions (aka loss or risk functions), which quantify the desirability of outcomes given management objectives, assigning higher values to preferred outcomes and lower values to less desirable alternatives. By maximizing expected utility, it identifies the optimal action among available actions. It thus yields explicit decision rules that link knowledge to action in a transparent and reproducible way.

Decision analysis has a long history in natural resource and wildlife management, notably through structured decision making and adaptive management (\citeproc{ref-Williams2002}{Williams et al. 2002}, \citeproc{ref-StructuredDecisionMaking2020}{Runge et al. 2020}). Within ecological statistics, early contributions strongly advocated the use of Bayesian approaches to integrate ecological inference with formal decision-making (e.g. \citeproc{ref-dorazio2003}{Dorazio and Johnson 2003}, \citeproc{ref-williams2016}{Williams and Hooten 2016}). Numerous applications in wildlife and natural resource management illustrate how uncertainty about ecological systems can be incorporated into the evaluation of management alternatives (e.g. \citeproc{ref-Rout2009}{Rout et al. 2009}, \citeproc{ref-Smith2011}{Smith et al. 2011}, \citeproc{ref-Moore2012}{Moore and Runge 2012}, \citeproc{ref-Green2015}{Green and Bailey 2015}, \citeproc{ref-Sells2016}{Sells et al. 2016}). These applications implement principles that closely align with Bayesian decision theory, without necessarily using that terminology or referring to the statistical decision-theory literature. This suggests that the gap is not simply a lack of decision-theoretic approaches in ecology, but also a disconnect between Bayesian statistical inference and the well-established practice of decision analysis. Bayesian decision theory provides an explicit framework for reconnecting these two components, by showing how posterior uncertainty can be propagated directly into the evaluation and optimization of management actions.

Strengthening the connection between inference and decision is particularly important in wildlife management, where decisions carry high stakes and are often contested. Managers must weigh uncertain ecological outcomes against economic costs, conservation goals and social acceptability (\citeproc{ref-converse2013matter}{Converse et al. 2013}). In such contexts, making assumptions explicit and quantifying trade-offs is not only scientifically desirable but also essential for transparency and legitimacy.

Here, we argue that Bayesian decision theory provides a practical framework to strengthen the connection between Bayesian inference and decision analysis and move ecology towards a more decision-oriented science. Our aim is not to develop new theory, but to show how existing concepts can be implemented in ecological applications. We first introduce the key ideas using a simple and intuitive example. We then propose a general workflow that can be applied with standard Bayesian tools. Finally, we illustrate the approach with two case studies in wildlife management, focusing on large carnivore management and invasive species control. These examples show how hierarchical state-space models can be directly linked to quantified trade-offs and optimized decisions. In both case studies we use real data and address real management problems, but we developed the decision analyses for pedagogical purposes rather than in collaboration with decision-makers for direct operational implementation. Their formulation nevertheless draws on our experience with large carnivore and invasive rodent management and was informed by the management objectives and trade-offs we encountered in these contexts.

\section{Bayesian decision theory}\label{bayesian-decision-theory}

\subsection{A simple introductory example}\label{a-simple-introductory-example}

We begin with a simple example designed to build intuition. Consider a computer whose condition is uncertain. Suppose that, based on available data, you can assign probabilities to two possible states: good (0.4) or poor (0.6).

You must now choose between two possible actions: either you repair your computer, or you do nothing. Each of these actions leads to a different outcome depending on the true state. To compare these outcomes, we assign a utility to each, reflecting its desirability. Suppose that a fully functioning computer provides a benefit of \texteuro{}30. We assume, for simplicity, that repairing the computer always restores it to full functionality and costs \texteuro{}10, regardless of its initial condition. The net benefit of repair is therefore \texteuro{}20 in either state. If you choose not to repair, a computer in good condition provides the full benefit (\texteuro{}30), whereas a computer in poor condition provides no benefit (\texteuro{}0). We define utility as this net monetary benefit. The utilities are summarized in Table \ref{tab:utility-computer}.

\begin{table}[ht]
\centering
\begin{tabular}{lcc}
\hline
 & \multicolumn{2}{c}{\textbf{Actions}} \\
\textbf{Condition} & Repair & Do nothing \\
\hline
{Good} & {\texteuro{}20} & {\texteuro{}30} \\
{Poor} & {\texteuro{}20} & {\texteuro{}0} \\
\hline
\end{tabular}
\caption{Utility associated with two possible actions (repair or do nothing) under different states of the system (computer condition). Utility is expressed as net monetary benefit, assuming that a fully functioning computer provides a benefit of \texteuro{}30 and that repair costs \texteuro{}10 and restores full functionality. Higher values indicate more desirable outcomes.}
\label{tab:utility-computer}
\end{table}

The values in Table \ref{tab:utility-computer} provide a deliberately simple and transparent utility scale. In practice, a utility function may combine multiple objectives, with their relative importance elicited from stakeholders or derived from management objectives.

Because the true state of the computer is unknown, we compute an average utility, weighted by the probabilities of the different states. This is the expected utility. We compute this for each action. For the action ``repair'', we obtain: \(0.4 \times 20 + 0.6 \times 20 = 20\), while for the action ``do nothing'', we obtain: \(0.4 \times 30 + 0.6 \times 0 = 12\). Therefore, repairing leads to an expected utility of \texteuro{}20, whereas doing nothing leads to an expected utility of \texteuro{}12. We therefore select the action that maximizes expected utility, which here is to repair the computer.

This example illustrates the key idea of Bayesian decision theory: decisions are made by averaging over uncertainty rather than relying on a single estimate.

\subsection{General framework}\label{general-framework}

The previous example can be generalized into a framework applicable to ecological decision-making. We define the core components of Bayesian decision theory. First, we define the state of nature, denoted by \(\theta\), which represents the unknown state of the system (e.g., population size, growth rate, or system condition). Given data \(y\), uncertainty about \(\theta\) is summarized by the posterior distribution \(p(\theta \mid y)\), representing current knowledge about the system. Second, we define a set of possible actions: \(\mathcal{A} = \{a_1, a_2, \dots\}\). These represent the decision alternatives available to the manager. Third, we specify a utility function: \(U(\theta, a)\) which quantifies the consequence of taking action \(a\) when the true state of the system is \(\theta\). It encodes objectives and trade-offs (e.g., benefits vs.~costs, ecological vs.~social outcomes), as well as the values of the decision-maker. Because the true state \(\theta\) is unknown, we evaluate each action by averaging its utility over all possible states, weighted by their probabilities. This leads to the expected utility:

\begin{equation}
\mathbb{E}[U(a)] = \int_{\Theta} U(\theta, a)\, p(\theta \mid y)\, d\theta
\label{eq:expected-utility}
\end{equation}

The expected utility in Equation \eqref{eq:expected-utility} represents the average performance of action \(a\), taking into account uncertainty. An important distinction needs to be made between linear and nonlinear utility functions. When utility is linear in an uncertain outcome, expected utility is equivalent to utility evaluated at the expected outcome, so carrying the full posterior distribution into the decision calculation may not affect the result. When utility is nonlinear, however, these quantities generally differ, and the distribution of possible outcomes becomes relevant. This occurs, for example, when consequences involve thresholds, asymmetric costs, or attitudes toward risk. More generally, the posterior distribution can be used to quantify risk explicitly, such as the probability of an undesirable outcome. This is illustrated in the wolf case study below. This also provides a connection with risk analysis, where decisions depend not only on expected outcomes but also on the probabilities and consequences of undesirable outcomes.

The optimal decision is the action that maximizes expected utility:

\begin{equation}
a^* = \arg\max_{a \in \mathcal{A}} \mathbb{E}[U(a)].
\label{eq:optimal-decision}
\end{equation}

This framework can be directly illustrated using the computer example introduced above. In that case, the state of nature corresponds to the condition of the computer: \(\theta \in \Theta = \{\text{good}, \text{poor}\}\) and uncertainty about this state is described by the distribution: \(p(\theta \mid y) = \{0.4, 0.6\}\). The set of possible actions is: \(\mathcal{A} = \{\text{repair}, \text{do nothing}\}\). The utility function \(U(\theta, a)\) is defined by Table \ref{tab:utility-computer}, which assigns a value to each combination of state and action. In this discrete setting, expected utility in Equation \eqref{eq:expected-utility} is computed as a weighted sum over all possible states: \(\mathbb{E}[U(a)] = \sum_{\theta \in \Theta} U(\theta, a)\, p(\theta \mid y)\). Applying this to the two actions gives: \(\mathbb{E}[U(\text{repair})] = 0.4 \times 20 + 0.6 \times 20 = 20\) and \(\mathbb{E}[U(\text{do nothing})] = 0.4 \times 30 + 0.6 \times 0 = 12\). The optimal decision is therefore: \(a^* = \arg\max_{a \in \mathcal{A}} \mathbb{E}[U(a)]\) which, in this case, corresponds to repairing the computer.

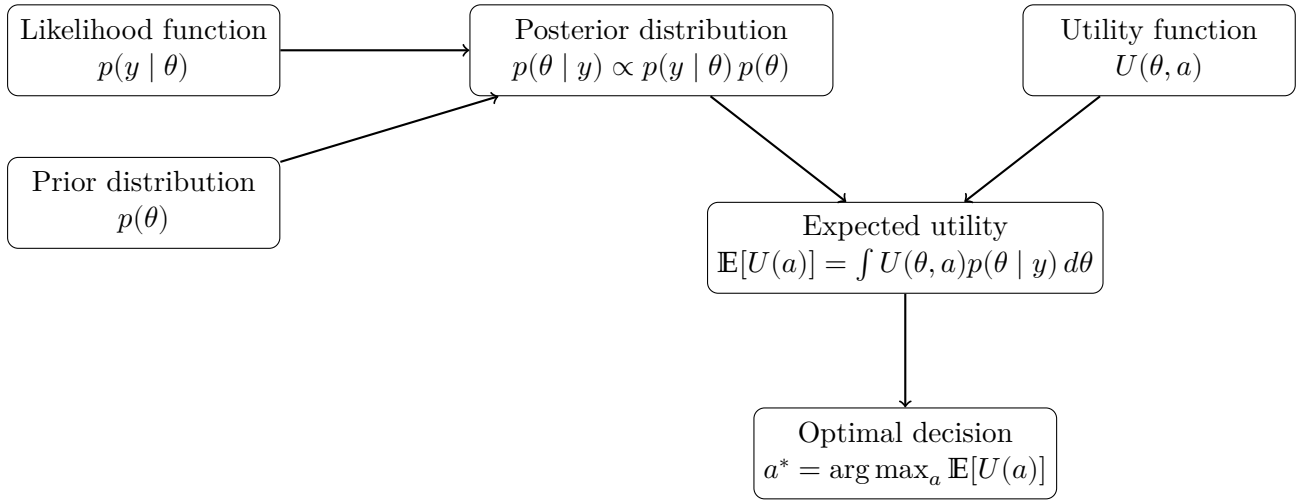
\begin{figure}[ht]
\centering
\begin{tikzpicture}[
  node distance=2cm,
  box/.style={rectangle, rounded corners, draw, align=center, minimum width=3.6cm, minimum height=1.2cm},
  arrow/.style={->, thick}
]

\node[box] (prior)
{Prior distribution\\ $p(\theta)$};

\node[box, above=0.8cm of prior] (likelihood)
{Likelihood function\\ $p(y \mid \theta)$};

\node[box, minimum width=4.8cm, right=2.5cm of likelihood] (posterior)
{Posterior distribution\\ 
$p(\theta \mid y) \propto p(y \mid \theta)\, p(\theta)$};

\node[box, right=2.5cm of posterior] (utility)
{Utility function\\ $U(\theta, a)$};

\node[box, minimum width=4.8cm, below=2cm of $(posterior)!0.5!(utility)$] (eu)
{Expected utility\\ 
$\mathbb{E}[U(a)] = \int U(\theta,a)p(\theta \mid y)\,d\theta$};

\node[box, below=1.5cm of eu] (decision)
{Optimal decision\\ $a^* = \arg\max_a \mathbb{E}[U(a)]$};

\draw[arrow] (prior) -- (posterior);
\draw[arrow] (likelihood) -- (posterior);

\draw[arrow] (posterior) -- (eu);
\draw[arrow] (utility) -- (eu);

\draw[arrow] (eu) -- (decision);

\end{tikzpicture}
\caption{Bayesian decision theory workflow. A statistical model links data $y$ to ecological processes through a likelihood and a prior with parameter $\theta$. These are combined to derive the posterior distribution $p(\theta \mid y) \propto p(y \mid \theta)p(\theta)$. A set of candidate actions $a$ is defined. A utility function $U(\theta,a)$ specifies the consequences of each action. Expected utility $\mathbb{E}[U(a)]$ is computed by averaging over the posterior distribution. The optimal decision $a^*$ is the action that maximizes expected utility.}
\label{fig:workflow-bdt}
\end{figure}

This framework naturally leads to a practical workflow described in Figure \ref{fig:workflow-bdt}, which provides a direct link between statistical inference and decision-making.

\subsection{Practical implementation}\label{practical-implementation}

In real ecological applications, the expected utility cannot be computed analytically. This is because it involves integrating over the posterior distribution of \(\theta\), which is often high-dimensional, making the integral intractable. However, Bayesian inference provides a natural solution: posterior distributions can be approximated using Monte Carlo sampling. If we draw \(\theta^{(1)}, \dots, \theta^{(N)} \sim p(\theta \mid y)\), the expected utility in Equation \eqref{eq:expected-utility} can be approximated as:

\begin{equation}
\mathbb{E}[U(a)] \approx \frac{1}{N} \sum_{n=1}^N U(\theta^{(n)}, a)
\label{eq:approx-exputil}
\end{equation}

In other words, the expected utility is estimated by averaging the utility over a set of draws from the posterior distribution. This approximation is straightforward to implement and fully compatible with standard Bayesian tools. The optimal decision in Equation \eqref{eq:optimal-decision} is then obtained by maximizing this Monte Carlo estimate:

\begin{equation}
a^* \approx \arg\max_{a \in \mathcal{A}} \frac{1}{N} \sum_{n=1}^N U(\theta^{(n)}, a)
\label{eq:approx-optimdec}
\end{equation}

This makes Bayesian decision theory both conceptually simple and practically accessible: once posterior samples are available, decision-making reduces to a simple post-processing step.

\section{Case study 1: Wolf management in France}\label{case-study-1-wolf-management-in-france}

Wolves recolonized France in the 1990s, leading to increasing conflicts with livestock farmers. Each year, authorities set legal shooting harvests, typically expressed as a fixed proportion of the estimated population size. This context provides a natural setting for Bayesian decision theory. Population dynamics are uncertain, with variable growth and imperfect knowledge of abundance, yet decisions must be made annually. The problem also involves clear trade-offs between reducing damage and maintaining a viable population. Management therefore aims to balance costs and benefits while keeping population size above a critical threshold.

\subsection{Data and population model}\label{data-and-population-model}

We rely on annual estimates of wolf abundance obtained from capture-recapture analyses (\citeproc{ref-cubaynesImportanceAccountingDetection2010}{Cubaynes et al. 2010}, \citeproc{ref-milleretEstimatingWolfPopulation2026}{Milleret et al. 2026}), along with the number of wolves removed each year Figure \ref{fig:fig-wolf-all}A.

These estimates are derived from a long-term national monitoring program based on non-invasive genetic sampling conducted during winter. Samples (e.g.~scats, hairs) are collected across the distribution range and genotyped to identify individuals (\citeproc{ref-pirogStandardizationHighQualityMethodological2025}{Pirog et al. 2025}, \citeproc{ref-milleretEstimatingWolfPopulation2026}{Milleret et al. 2026}).

Importantly, these abundances are not raw counts but statistical estimates. We treat them as noisy observations of a latent population size. Observation uncertainty is directly informed by the reported confidence intervals.

The observation model links these capture-recapture estimate \(\hat{N}_t\) to the true latent population size in year \(t\) denoted \(N_t\). Because abundance estimates are positive and uncertainty is multiplicative, we model them on the log scale:

\begin{equation}
\log(\hat{N}_t) \sim \mathcal{N}\left(\log(N_t), \sigma_{obs,t}^2\right).
\label{eq:obs-wolf}
\end{equation}

The observation standard deviation \(\sigma_{obs,t}\) is derived from the reported 95\% confidence interval \([C_{t,lo},C_{t,hi}]\):

\begin{equation}
\sigma_{obs,t} = \frac{\log(C_{t,hi})-\log(C_{t,lo})}{2 \times 1.96}.
\label{eq:sdobs-wolf}
\end{equation}

Let \(H_t\) denote the number of wolves removed during year \(t\). We model population dynamics as a stochastic multiplicative process with removals:

\begin{equation}
N_t \sim \text{LogNormal}\left( \log\left[\lambda (N_{t-1}-H_{t-1})\right], \sigma_{proc}^2 \right).
\label{eq:process-wolf}
\end{equation}

Here, \(\lambda\) is the mean growth rate, while \(\sigma_{proc}\) captures interannual variability. The term \(N_{t-1}-H_{t-1}\) represents the population remaining after removals, before population growth occurs. Density dependence and resource limitation could be incorporated using a theta-logistic population model (e.g. \citeproc{ref-Regehr2021}{Regehr et al. 2021}), although estimating the associated carrying capacity can be challenging, particularly for expanding populations (see also \citeproc{ref-Clark2010}{Clark et al. 2010} regarding limitations of this model).

The full vector of unknown quantities is therefore \(\theta = \left(\lambda, \sigma_{proc}, N_1,\dots,N_T \right)\). We assigned weakly informative priors to all model parameters. The model was fitted in \texttt{NIMBLE} (\citeproc{ref-devalpineProgrammingModelsWriting2017a}{{de Valpine et al.} 2017}). We ran two MCMC chains (20,000 iterations, 2,000 burn-in) and assessed convergence using standard diagnostics (R-hat, effective sample size, trace plots).

\subsection{Utility function}\label{utility-function}

If \(T\) is the current year, management actions correspond to future removals \(a\), defined as the number of wolves that could be removed in the next year \(T+1\). These actions play the same role as ``repair'' or ``do nothing'' in the introductory example, but here they correspond to management decision alternatives.

We now formalize the trade-off faced by managers. On the one hand, increasing removals can be beneficial, for example by reducing livestock damage or alleviating social conflicts. On the other hand, excessive removals increase the risk of compromising the favourable conservation status of the species. We illustrate this trade-off using the following utility function:

\begin{equation}
U(a) = (b - c)a - \alpha \, P(N_{T+1} < N_{min})
\label{eq:utility-wolf}
\end{equation}

The first term represents the net benefit of removals: \(b\) is the benefit per individual removed (e.g., reduced damage or increased acceptability), \(c\) is the associated cost (e.g., ethical, political, or economic), and their difference \((b - c)\) defines the net gain per individual. The second term represents a penalty associated with risk. The quantity \(P(N_{T+1} < N_{min})\) is the probability that the population falls below a critical threshold \(N_{min}\), and \(\alpha\) determines how strongly this risk is penalized. The parameter \(\alpha\) determines how strongly this risk is penalized and should be interpreted relative to the scale of the net benefits of removal. For example, \(\alpha = 0\) implies no penalty for population risk, whereas with \(\alpha = 100\), a probability of 0.5 of falling below \(N_{min}\) results in a penalty of 50 utility units. In summary, the first term captures the benefits and costs of harvesting, while the second term penalizes the ecological risk associated with low abundance.

Both the form of the utility function and the values assigned to parameters such as \(\alpha\) and \(N_{min}\) reflect management preferences rather than quantities estimated from the ecological data and should ideally be defined through engagement with the decision-makers. Here, we use illustrative values and explore their influence through sensitivity analysis.

A key point is that this utility depends on a probability, and therefore on the distribution of the future population size. This distribution itself depends both on the action \(a\) and on uncertainty in the model parameters and current population size.

\subsection{Optimal decision}\label{optimal-decision}

To compute expected utility, we rely on posterior samples obtained via MCMC. For each draw \((\lambda^{(n)}, \sigma_{proc}^{(n)}, N_T^{(n)})\), we simulate a possible future population size \(N_{T+1}^{(n)}(a)\) under a given number of wolves to be harvested \(a\). In doing so, uncertainty in abundance, process variability, and the effect of harvest is naturally propagated into the decision analysis. To reduce Monte Carlo variability, multiple process realisations (500 times here) are simulated for each posterior draw, ensuring that both parameter uncertainty and process variability are adequately represented in the expected utility. We then approximate the probability \(P(N_{T+1} < N_{min})\) as the proportion of simulated outcomes falling below the threshold. This allows us to compute the expected utility \(\mathbb{E}[U(a)]\) for each candidate harvest as in Equation \eqref{eq:approx-exputil}.

\subsection{Results}\label{results}

\begin{figure}
\includegraphics[width=1\linewidth]{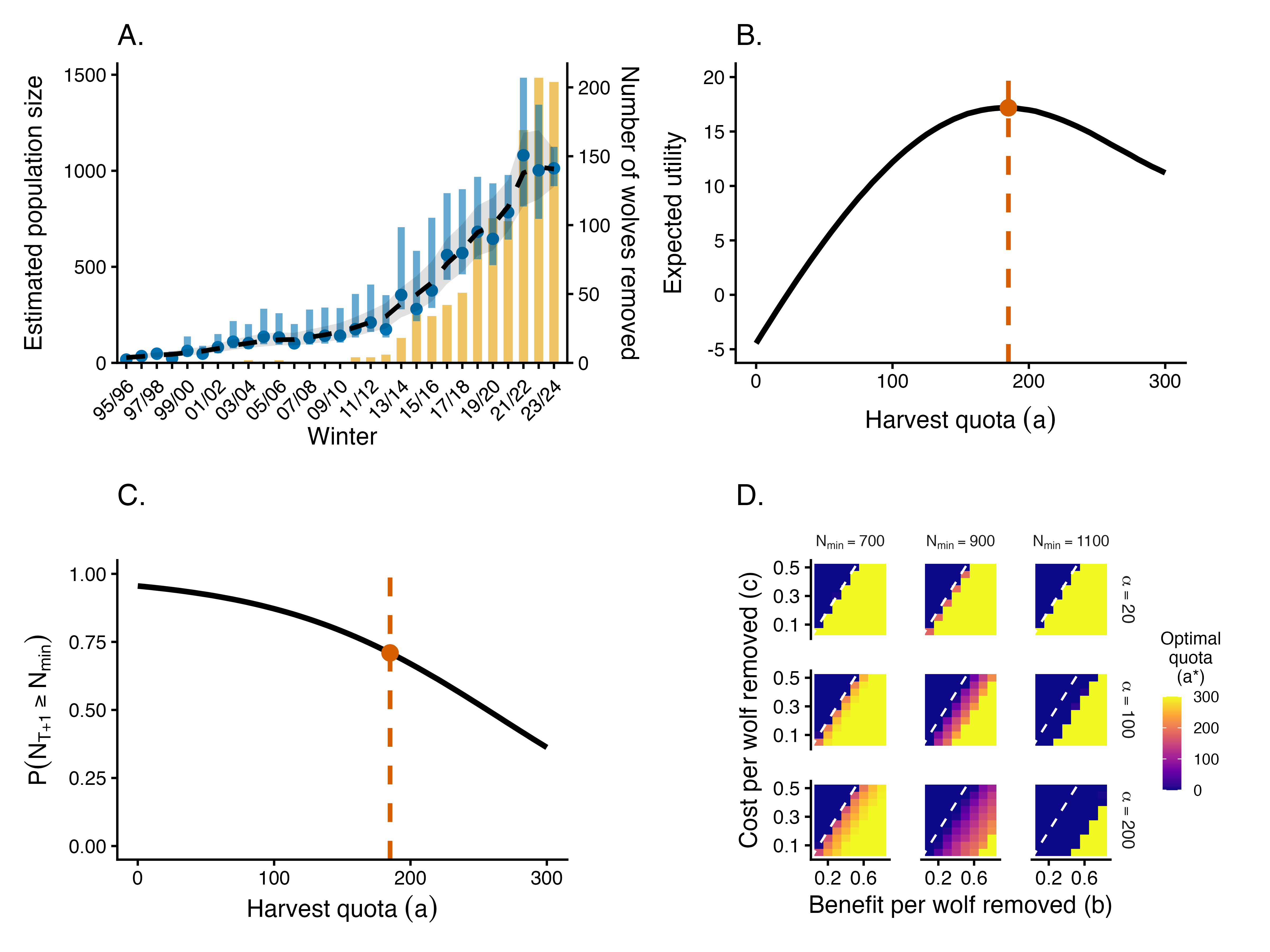} \caption{Bayesian decision framework for wolf management in France. (A) Temporal dynamics of wolf population size and removals. Blue points and vertical lines show capture-recapture estimates of population size with associated uncertainty, orange bars represent the number of wolves removed each year, and the dashed black line with shaded area corresponds to posterior mean and 95\% credible intervals of the latent population size. (B) Expected utility as a function of the harvest, showing an intermediate optimum resulting from the trade-off between benefits of removals and ecological risk. The vertical orange line indicates the optimal harvest. (C) Probability that the population remains above the minimum threshold ($N_{min} = 900$) as a function of the harvest, illustrating the increase in extinction risk with higher removals. The vertical orange line indicates the optimal harvest identified in panel (B). (D) Optimal harvest across combinations of management preferences, defined by the benefit ($b$) and cost ($c$) of removals, for different levels of risk aversion (\(\alpha\)) and minimum population threshold ($N_{min}$). The diagonal line separates situations where costs exceed benefits (no harvest) from those where removals are beneficial.}\label{fig:fig-wolf-all}
\end{figure}

The model reproduces the main temporal trends in abundance and removals (Figure \ref{fig:fig-wolf-all}A), with most observations falling within the 95\% credible intervals, indicating an adequate basis for decision analysis.

To illustrate the decision framework, we use baseline management preferences: benefit per individual removed \(b=0.4\), cost \(c=0.15\), risk aversion \(\alpha=100\), and minimum acceptable population size \(N_{\text{min}}=900\). These values are illustrative; sensitivity is explored in Figure \ref{fig:fig-wolf-all}D.

Expected utility increases at low harvest levels and then declines as ecological risk rises (Figure \ref{fig:fig-wolf-all}B). The maximum defines the optimal harvest \(a^*\), balancing benefits, costs, and risk. This trade-off is clearer in Figure \ref{fig:fig-wolf-all}C: the probability of remaining above the threshold decreases with harvest, indicating increasing risk. The optimal harvest therefore accepts a non-zero level of risk.

Sensitivity analysis (Figure \ref{fig:fig-wolf-all}D) shows that higher risk aversion (\(\alpha\)) or stricter thresholds (\(N_{min}\)) lead to more conservative strategies. Such sensitivity analyses are particularly important when management preferences and the relative weights assigned to different objectives are uncertain, as they show whether the decision is robust to alternative values. Thus, optimal harvest depends not only on ecological dynamics but also on how trade-offs and risk are valued.

\section{Case study 2: Muskrat management in the Netherlands}\label{case-study-2-muskrat-management-in-the-netherlands}

We now extend the Bayesian decision-theoretic framework to a more complex setting: the management of invasive muskrat populations in the Netherlands (\citeproc{ref-vanloon2017}{Loon et al. 2017}). Muskrat is an invasive species whose burrowing compromises dykes and poses a public safety risk. Control has been implemented since its introduction in 1941.

Unlike the wolf example, where the decision consists in choosing a single harvest at the population level, muskrat management is inherently spatial. Control is implemented locally through trapping, and managers must decide how to allocate a limited control effort across space, rather than simply how much to remove in total. Population dynamics are spatially structured, detection is imperfect and depends on effort, and management actions correspond to effort allocation across provinces. As in the wolf case, decisions must be made under uncertainty and involve trade-offs, but the decision space is now multidimensional.

\subsection{Data and population model}\label{data-and-population-model-1}

To investigate the relation between catch and effort and enhance prediction models, a large randomized controlled experiment was conducted from 2013 to 2016 (\citeproc{ref-bosLargeScaleExperimentEvaluate2020a}{{Bos et al.} 2020}). The experimental units were 5 × 5 km grid cells (hereafter ``sites'', corresponding to atlas squares in the Dutch national coordinate system), covering a total of 2202 locations across the mainland. Data were aggregated at the seasonal level (four seasons per year), resulting in 13 successive time steps from winter 2012/2013 to winter 2015/2016.

We rely on these large-scale monitoring data (Figure \ref{fig:fig-muskrat-all}A), which include, for each site and season, the number of muskrats removed (``catch'') and the associated trapping effort, measured as the total number of hours spent trapping. These site-level data are used to fit the population model, while management decisions are formulated at a higher spatial level by allocating effort across provinces, which aggregate multiple sites. This distinction between sites (where the ecological process is modelled) and provinces (where decisions are applied) is central to the spatial decision framework.

At each site \(i\) and time \(t\), we observe the number of individuals removed \(y_{i,t}\), while trapping effort is considered and modelled at the province level. The true abundance \(N_{i,t}\) is latent. The observation model assumes that removals arise from a binomial process:

\begin{equation}
y_{i,t} \sim \text{Binomial}(N_{i,t}, p_{i,t}),
\label{eq:obs-muskrat}
\end{equation}

where capture probability depends on effort through a complementary log-log link

\begin{equation}
\text{cloglog}(p_{i,t}) = \alpha_0 + \alpha_{\text{Eff}} \log(Eff_{i,t}^{prov} + 1).
\label{eq:obs-cloglog}
\end{equation}

where \(Eff_{i,t}^{prov}\) denotes the average trapping effort in the province to which site \(i\) belongs at time \(t\) and the \(\alpha\)'s are parameters to be estimated. This formulation links site-level removals to trapping intensity summarized at the provincial level and ensures that capture probability increases with effort while remaining bounded between 0 and 1.

Population dynamics are modeled using a density-dependent process. After removal, the remaining population is \(R_{i,t} = N_{i,t} - y_{i,t}\). We then model the next abundance as:

\begin{equation}
N_{i,t+1} \sim \text{Poisson}(\mu_{i,t}),
\label{eq:process-muskrat}
\end{equation}

where

\begin{equation}
\log \mu_{i,t} =
\log(R_{i,t}) +
r +
\beta_{\text{density}} N_{i,t} +
\beta_{\text{neighb}} \, \text{neighb}_{i,t},
\label{eq:meanprocess-muskrat}
\end{equation}

where \(r\) is the density-independent component of population growth, \(\beta_{\text{density}}\) captures density dependence, and \(\text{neighb}_{i,t}\) represents a connectivity covariate defined as the average catch per unit effort in the eight neighboring atlas squares surrounding site \(i\), with \(\beta_{\text{neighb}}\) quantifying its effect.

The full set of unknown quantities is therefore \(\theta = \left(\alpha_0, \alpha_{\text{Eff}}, r, \beta_{\text{density}}, \beta_{\text{neighb}}, N_{i,t}\right)\). We assigned weakly informative priors to all model parameters. We ran two MCMC chains for 50,000 iterations each and discarded the first 15,000 iterations as burn-in.

\subsection{Utility function}\label{utility-function-1}

Here, management actions correspond to allocations of trapping effort across provinces. Let \(\mathbf{a} = (a_1, \dots, a_P)\) denote the effort allocated to each province, with \(P\) the number of provinces. Unlike the wolf case, the action is now a vector rather than a scalar.

We consider a simple utility function that captures the trade-off between the cost of control and the resulting population size:

\begin{equation}
U(\mathbf{a}) = - c \sum_{p=1}^{P} a_p - \gamma \sum_{i=1}^{I} N_{i,T+1}(\mathbf{a}).
\label{eq:utility-muskrat}
\end{equation}

The first term represents the total cost of control, with \(c\) denoting the cost per unit of trapping effort and \(\sum_p a_p\) the total effort deployed across provinces. The second term represents the cost associated with residual population size, aggregated across sites. The parameter \(\gamma\) controls the relative importance of population reduction versus effort cost.

As in the wolf example, both the form of the utility function and the values assigned to parameters such as \(c\) and \(\gamma\) reflect management preferences rather than quantities estimated from the ecological data and should ideally be defined through engagement with the decision-makers. Here, we use illustrative values and explore their influence through sensitivity analysis.

This formulation differs from the wolf case in two key ways. First, the objective is not to maintain the population above a threshold, but rather to minimize overall abundance, reflecting the invasive status of the species. Second, the decision involves distributing effort across space, rather than choosing a single removal level.

As before, the utility depends on the distribution of future abundance, which itself depends on both the action \(\mathbf{a}\) and uncertainty in the model parameters and latent states.

\subsection{Optimal decision}\label{optimal-decision-1}

Expected utility is computed by integrating over posterior uncertainty. For each posterior draw \(\theta^{(n)}\), we calculate the expected future abundance \(N_{i,T+1}(\mathbf{a})\) under a given effort allocation \(\mathbf{a}\) using the conditional expectation implied by the population model. These expectations are then averaged across posterior draws, thereby propagating uncertainty in model parameters and latent population states into the decision analysis. Because the utility function is linear in future abundance, this calculation does not require additional simulation of process variability.

We consider two related decision problems. First, we evaluate a scalar action corresponding to the same mean trapping effort per site applied in every province, by exploring a range of candidate effort values. For each candidate effort, we calculate its expected utility and identify the effort level that maximizes it. We also examine the sensitivity of this optimal effort to the relative cost of effort (\(c\)) and penalty on residual abundance (\(\gamma\)).

Second, we consider a fixed total effort budget and optimize its allocation across provinces. Here, \(a_p\) represents the total effort allocated to province \(p\); because provinces contain different numbers of sites, this total is converted to mean effort per site within each province before calculating capture probability. We identify the allocation that maximizes expected utility subject to the constraint \(\sum_{p=1}^{P} a_p = B\), where \(B\) is the total available effort. We repeat this optimization across a range of total budgets to examine how the optimal spatial allocation changes with effort availability.

\subsection{Results}\label{results-1}

\begin{figure}
\includegraphics[width=1\linewidth]{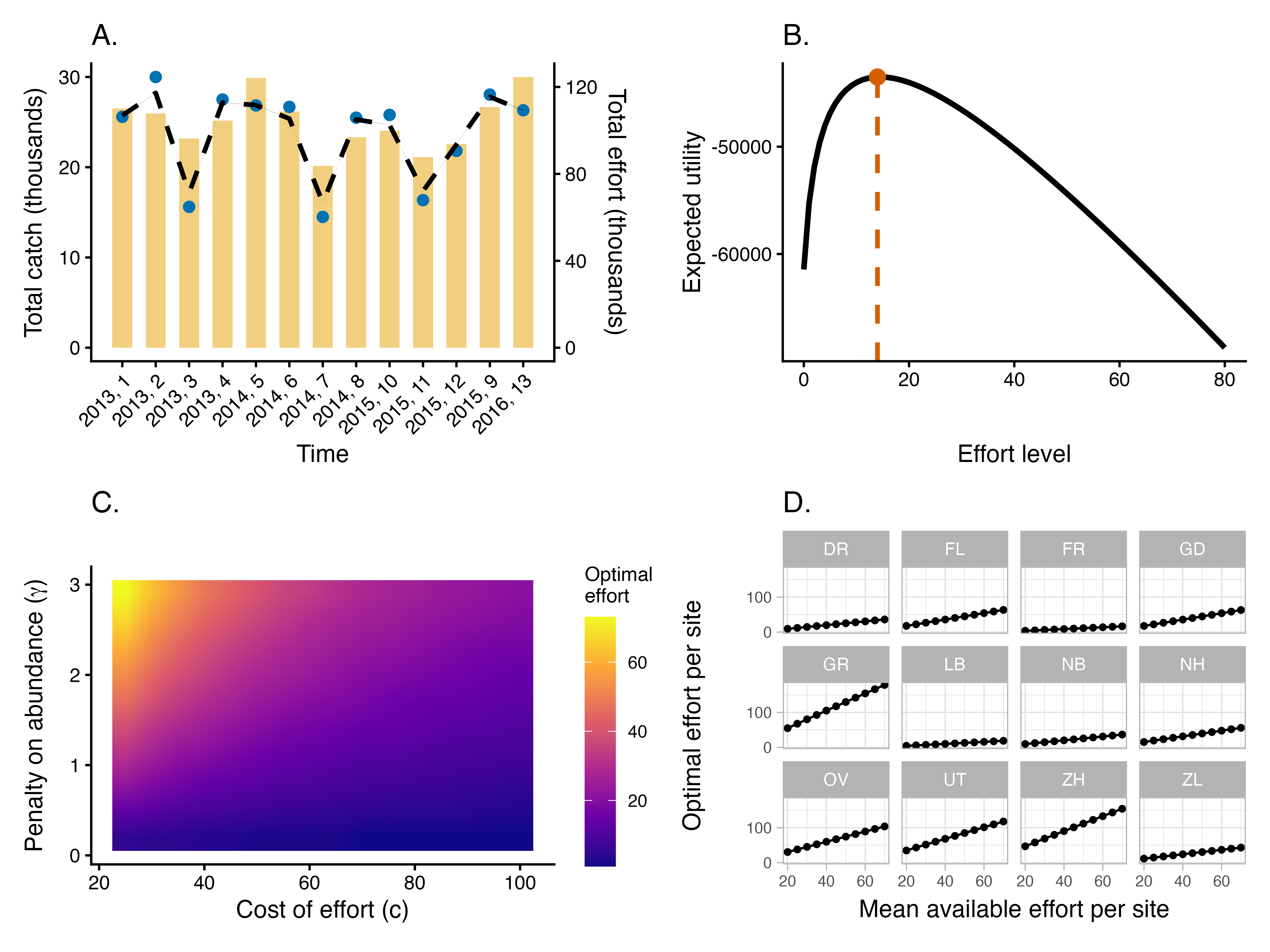} \caption{Bayesian decision framework for muskrat management in the Netherlands. (A) Posterior predictions of total removals over time. Observed catches (blue points) are compared with model predictions (black dashed line) and their 95\% credible intervals (shaded area). Trapping effort (orange bars) is shown on a secondary axis after rescaling. (B) Expected utility as a function of uniform mean trapping effort per site applied across all provinces. The vertical dashed line indicates the optimal effort level that maximizes expected utility. (C) Sensitivity of the optimal effort to management preferences, showing how the trade-off between control cost ($c$) and penalty on residual abundance (\(\gamma\)) influences the decision. (D) Optimal spatial allocation of trapping effort across provinces as the total available effort increases. The x-axis represents mean available effort per site across all provinces, while each panel shows the corresponding optimal mean trapping effort per site within a province. Differences among provinces illustrate how the optimal allocation of a fixed total effort budget varies spatially.}\label{fig:fig-muskrat-all}
\end{figure}

Observed catches and posterior predictions show good agreement (Figure \ref{fig:fig-muskrat-all}A), indicating that the model captures both the magnitude and temporal variation in removals despite variability in effort.

We use baseline preferences with effort cost \(c=50\) and penalty on residual abundance \(\gamma=1\). Expected utility increases sharply at low effort levels, reaches a maximum at an effort level of \(14\), and then decreases as the marginal benefit of additional control is outweighed by its increasing cost (Figure \ref{fig:fig-muskrat-all}B). Optimal effort increases with the penalty on abundance and decreases with control cost (Figure \ref{fig:fig-muskrat-all}C), reflecting a continuous trade-off between control costs and population reduction.

We then consider a spatial decision problem: instead of applying the same effort in all provinces, we allocate a fixed total effort budget \(B\) among provinces with \(\sum_{p=1}^{P} a_p = B\). The optimal allocation is strongly spatially heterogeneous (Figure \ref{fig:fig-muskrat-all}D). As the total available effort increases, optimal trapping effort increases in all provinces, but at markedly different rates. In particular, Groningen (GR) and South Holland (ZH) consistently receive the highest effort per site, followed by Utrecht (UT) and Overijssel (OV), whereas substantially less effort is allocated to provinces such as Friesland (FR) and Limburg (LB). This spatial differentiation reflects differences among provinces in the expected marginal benefit of additional trapping effort, given the estimated population states, spatial dynamics, and response of capture probability to effort.

\section{Conclusion}\label{conclusion}

The core idea of Bayesian decision theory is simple. A statistical model yields a posterior distribution that represents uncertainty, each action generates a distribution of outcomes, and management objectives are formalized through a utility function encoding trade-offs. The optimal decision is then obtained by selecting the action that maximizes expected utility. Decision-making thus becomes a natural extension of inference. Bayesian decision theory therefore provides a direct link between what we estimate and what we decide (\citeproc{ref-berger1985statistical}{Berger 1985}).

The two case studies illustrate the flexibility of the framework, from a single scalar decision to spatially structured strategies. In the wolf example, the decision consisted in choosing a single harvest under uncertainty, highlighting the role of risk and trade-offs. In contrast, the muskrat example involves allocating effort across space under heterogeneous dynamics and imperfect detection. Together, they highlight the ability of Bayesian decision theory to accommodate increasing realism without changing its conceptual foundation.

More broadly, several messages emerge. First, decision-making is not purely technical. While models quantify uncertainty, defining the utility function---what counts as a cost or a benefit, and how these are weighted---inevitably involves social, political, and ethical considerations. Bayesian decision theory does not itself provide a process for framing the decision or eliciting objectives and values; these steps require careful engagement with decision-makers and stakeholders and are extensively addressed within structured decision making (\citeproc{ref-Williams2002}{Williams et al. 2002}, \citeproc{ref-StructuredDecisionMaking2020}{Runge et al. 2020}). The utility function formalizes management objectives by encoding how outcomes are valued, enabling explicit comparison of actions under uncertainty.

The relative weights assigned to different objectives can be informed by stakeholder elicitation, management targets, economic valuation, or other approaches for translating preferences into quantitative values. When multiple objectives are expressed in different units or on very different scales (e.g.~monetary costs and population outcomes), care must be taken when combining them within a single utility function. Outcomes may therefore need to be rescaled or normalized before being weighted and combined. Utility functions may also be nonlinear, for example to represent decreasing marginal utility or attitudes toward risk. In repeated decision problems, temporal discounting can additionally be used to assign different values to present and future outcomes. For detailed guidance on constructing and eliciting utility functions, we refer readers to Williams et al. (\citeproc{ref-Williams2002}{2002}) and Runge et al. (\citeproc{ref-StructuredDecisionMaking2020}{2020}).

Because choices about the form of the utility function and how outcomes are valued are context-dependent, different stakeholders may reach different optimal decisions from the same data. Making these assumptions explicit promotes transparency and facilitates dialogue among stakeholders (\citeproc{ref-keeney1996value}{Keeney 1996}).

Second, models remain simplified representations of ecological systems. In our case studies, dynamics could be refined by incorporating additional processes such as density dependence or trophic interactions. In spatial systems, improvements could also better account for connectivity. Utility functions could also incorporate additional dimensions such as monitoring costs or ecosystem services. Although the framework is flexible, its effective implementation requires stronger integration across disciplines (\citeproc{ref-hemmingIntroductionDecisionScience2022a}{Hemming et al. 2022}).

Third, we deliberately consider one-shot decisions in our examples to focus on the connection between inference, utility and action. In practice, management decisions such as harvest are repeated, and each action can affect both system dynamics and future information. This naturally leads to dynamic frameworks such as adaptive management and Markov decision processes, where the goal is to identify optimal policies over time (e.g. \citeproc{ref-marescotComplexDecisionsMade2013}{Marescot et al. 2013}). When learning about uncertain system dynamics is itself part of the decision problem, this framework can be extended to partially observable Markov decision processes (\citeproc{ref-Chades2021}{Chadès et al. 2021}) and its Bayesian counterpart (\citeproc{ref-Ross2007}{Ross et al. 2007}). A further direction is to use Bayesian decision theory not only to choose actions, but also to determine what information is worth collecting. Value of Information (VOI) analyses quantify the benefit of reducing uncertainty for a particular decision or management objective (\citeproc{ref-runge2011uncertainty}{Runge et al. 2011}). In particular, the expected value of sample information (EVSI) provides a natural framework for designing monitoring programs by quantifying the expected improvement in decision performance from collecting additional, imperfect information.

Finally, wider adoption will also require accessible computational tools that connect Bayesian inference directly to utility calculations, optimization, and VOI. Just as software such as \texttt{WinBUGS} (\citeproc{ref-lunn2000winbugs}{Lunn et al. 2000}), \texttt{JAGS} (\citeproc{ref-jags_software}{Plummer 2003}), \texttt{Stan} (\citeproc{ref-Carpenter2017Stan}{Carpenter et al. 2017}) and \texttt{NIMBLE} (\citeproc{ref-deValpine2017}{Valpine et al. 2017}) helped make Bayesian inference widely accessible in ecology, developing comparable tools for decision analysis could lower the barrier from fitting models to using them to inform decisions. Such developments could help make decision-oriented modelling a more routine extension of Bayesian ecological analysis.

\section{References}\label{references}

\protect\phantomsection\label{refs}


\begin{CSLReferences}{1}{0}
\bibitem[\citeproctext]{ref-berger1985statistical}
Berger, J. O. 1985. Statistical decision theory and {B}ayesian analysis. Springer-Verlag, New York.

\bibitem[\citeproctext]{ref-bosLargeScaleExperimentEvaluate2020a}
{Bos, D., E. E. van Loon, E. Klop, and R. Ydenberg}. 2020. A {Large-Scale Experiment} to {Evaluate Control} of {Invasive Muskrats}. Wildlife Society Bulletin 44:314--322.

\bibitem[\citeproctext]{ref-Carpenter2017Stan}
Carpenter, B., A. Gelman, M. D. Hoffman, D. Lee, B. Goodrich, M. Betancourt, M. Brubaker, J. Guo, P. Li, and A. Riddell. 2017. \href{https://doi.org/10.18637/jss.v076.i01}{Stan: A probabilistic programming language}. Journal of Statistical Software 76:1--32.

\bibitem[\citeproctext]{ref-Chades2021}
Chadès, I., L. V. Pascal, S. Nicol, C. S. Fletcher, and J. Ferrer-Mestres. 2021. A primer on partially observable markov decision processes (POMDPs). Methods in Ecology and Evolution 12:2058--2072.

\bibitem[\citeproctext]{ref-Clark2010}
Clark, F., B. W. Brook, S. Delean, H. R. Akçakaya, and C. J. A. Bradshaw. 2010. The theta-logistic is unreliable for modelling most census data. Methods in Ecology and Evolution 1:253--262.

\bibitem[\citeproctext]{ref-converse2013matter}
Converse, S. J., C. T. Moore, M. J. Folk, and M. C. Runge. 2013. A matter of tradeoffs: Reintroduction as a multiple objective decision. Journal of Wildlife Management 77:1145--1156.

\bibitem[\citeproctext]{ref-cubaynesImportanceAccountingDetection2010}
Cubaynes, S., R. Pradel, R. Choquet, C. Duchamp, J.-M. Gaillard, J.-D. Lebreton, E. Marboutin, C. Miquel, A.-M. Reboulet, C. Poillot, P. Taberlet, and O. Gimenez. 2010. Importance of accounting for detection heterogeneity when estimating abundance: The case of {French} wolves. Conservation Biology 24:621--626.

\bibitem[\citeproctext]{ref-devalpineProgrammingModelsWriting2017a}
{de Valpine, P., D. Turek, C. J. Paciorek, C. Anderson-Bergman, D. T. Lang, and R. Bodik}. 2017. Programming {With Models}: {Writing Statistical Algorithms} for {General Model Structures With NIMBLE}. Journal of Computational and Graphical Statistics 26:403--413.

\bibitem[\citeproctext]{ref-dorazio2003}
Dorazio, R. M., and F. A. Johnson. 2003. Bayesian inference and decision theory - {A} framework for decision making in natural resource management. Ecological Applications 13:556--563.

\bibitem[\citeproctext]{ref-Green2015}
Green, A. W., and L. L. Bailey. 2015. Using bayesian population viability analysis to define relevant conservation objectives. Plos One 10:e0144786.

\bibitem[\citeproctext]{ref-hemmingIntroductionDecisionScience2022a}
Hemming, V., A. E. Camaclang, M. S. Adams, M. Burgman, K. Carbeck, J. Carwardine, I. Chadès, L. Chalifour, S. J. Converse, L. N. K. Davidson, G. E. Garrard, R. Finn, J. R. Fleri, J. Huard, H. J. Mayfield, E. M. Madden, I. Naujokaitis-Lewis, H. P. Possingham, L. Rumpff, M. C. Runge, D. Stewart, V. J. D. Tulloch, T. Walshe, and T. G. Martin. 2022. An introduction to decision science for conservation. Conservation Biology 36:e13868.

\bibitem[\citeproctext]{ref-hobbs2025}
Hobbs, N. T., and M. B. Hooten. 2025. {B}ayesian {M}odels: {A} {S}tatistical {P}rimer for {E}cologists, 2nd {E}dition. Princeton University Press.

\bibitem[\citeproctext]{ref-keeney1996value}
Keeney, R. L. 1996. Value-focused thinking: A path to creative decisionmaking. Harvard University Press.

\bibitem[\citeproctext]{ref-KR2020}
Kéry, M., and J. A. Royle. 2020. {A}pplied {H}ierarchical {M}odeling in {E}cology: {A}nalysis of {D}istribution, {A}bundance and {S}pecies {R}ichness in {R} and {BUGS}, {V}olume {T}wo: {D}ynamic and {A}dvanced {M}odels. Academic Press, London, UK.

\bibitem[\citeproctext]{ref-vanloon2017}
Loon, E. E. van, D. Bos, C. J. van Hellenberg Hubar, and R. C. Ydenberg. 2017. A historical perspective on the effects of trapping and controlling the muskrat (\emph{{O}ndatra zibethicus}) in the {N}etherlands. Pest Management Science 73:305--312.

\bibitem[\citeproctext]{ref-lunn2000winbugs}
Lunn, D. J., A. Thomas, N. Best, and D. Spiegelhalter. 2000. WinBUGS -- a bayesian modelling framework: Concepts, structure, and extensibility. Statistics and Computing 10:325--337.

\bibitem[\citeproctext]{ref-marescotComplexDecisionsMade2013}
Marescot, L., G. Chapron, I. Chadès, P. L. Fackler, C. Duchamp, E. Marboutin, and O. Gimenez. 2013. Complex decisions made simple: A primer on stochastic dynamic programming. Methods in Ecology and Evolution 4:872--884.

\bibitem[\citeproctext]{ref-milleretEstimatingWolfPopulation2026}
Milleret, C., C. Duchamp, S. Bauduin, C. Kaerle, A. Pirog, G. Queney, and O. Gimenez. 2026. Estimating wolf population size in {France} using non-invasive genetic sampling and spatial capture-recapture models. Biological Conservation 317:111772.

\bibitem[\citeproctext]{ref-Moore2012}
Moore, J. L., and M. C. Runge. 2012. Combining structured decision making and value-of-information analyses to identify robust management strategies. Conservation Biology 26:810--820.

\bibitem[\citeproctext]{ref-pirogStandardizationHighQualityMethodological2025}
Pirog, A., C. Duchamp, C. Kaerle, C. Dufaure de Citres, S. Rousselot, J. Lavarec, and G. Queney. 2025. Standardization of a {High-Quality Methodological Framework} for {Long-Term Genetic Monitoring} of the {French Wolf Population}. Ecology and Evolution 15:e71345.

\bibitem[\citeproctext]{ref-jags_software}
Plummer, M. 2003. JAGS: A program for analysis of bayesian graphical models using gibbs sampling.

\bibitem[\citeproctext]{ref-polaskyDecisionmakingGreatUncertainty2011}
Polasky, S., S. R. Carpenter, C. Folke, and B. Keeler. 2011. Decision-making under great uncertainty: Environmental management in an era of global change. Trends in Ecology \& Evolution 26:398--404.

\bibitem[\citeproctext]{ref-Regehr2021}
Regehr, E. V., M. Dyck, S. Iverson, D. S. Lee, N. J. Lunn, J. M. Northrup, M.-C. Richer, G. Szor, and M. C. Runge. 2021. Incorporating climate change in a harvest risk assessment for polar bears {Ursus maritimus} in southern hudson bay. Biological Conservation 258:109128.

\bibitem[\citeproctext]{ref-Ross2007}
Ross, S., B. Chaib-draa, and J. Pineau. 2007. Bayes-adaptive {POMDPs}. Pages 1225--1232 \emph{in} J. C. Platt, D. Koller, Y. Singer, and S. T. Roweis, editors. Advances in neural information processing systems 20. Curran Associates, Inc., Vancouver, British Columbia, Canada.

\bibitem[\citeproctext]{ref-Rout2009}
Rout, T. M., C. E. Hauser, and H. P. Possingham. 2009. Optimal adaptive management for the translocation of a threatened species. Ecological Applications 19:515--526.

\bibitem[\citeproctext]{ref-RD2008}
Royle, J. A., and R. M. Dorazio. 2008. {H}ierarchical {M}odeling and {I}nference in {E}cology: {T}he {A}nalysis of {D}ata from {P}opulations, {M}etapopulations and {C}ommunities. Academic Press. San Diego, California.

\bibitem[\citeproctext]{ref-runge2011uncertainty}
Runge, M. C., S. J. Converse, and J. E. Lyons. 2011. Which uncertainty? Using expert elicitation and expected value of information to design an adaptive program. Biological Conservation 144:1214--1223.

\bibitem[\citeproctext]{ref-StructuredDecisionMaking2020}
Runge, M. C., S. J. Converse, J. E. Lyons, and D. R. Smith (Eds.). 2020. Structured {Decision Making}: {Case Studies} in {Natural Resource Management}. Johns Hopkins University Press, Baltimore, Maryland.

\bibitem[\citeproctext]{ref-Sells2016}
Sells, S. N., M. S. Mitchell, V. L. Edwards, J. A. Gude, and N. J. Anderson. 2016. Structured decision making for managing pneumonia epizootics in bighorn sheep. The Journal of Wildlife Management 80:957--969.

\bibitem[\citeproctext]{ref-Smith2011}
Smith, D. H. V., S. J. Converse, K. W. Gibson, A. Moehrenschlager, W. A. Link, G. H. Olsen, and K. Maguire. 2011. Decision analysis for conservation breeding: Maximizing production for reintroduction of whooping cranes. The Journal of Wildlife Management 75:501--508.

\bibitem[\citeproctext]{ref-deValpine2017}
Valpine, P. de, D. Turek, C. J. Paciorek, C. Anderson-Bergman, D. Temple Lang, and R. Bodik. 2017. Programming with models: Writing statistical algorithms for general model structures with NIMBLE. Journal of Computational and Graphical Statistics 26:403--413.

\bibitem[\citeproctext]{ref-Williams2002}
Williams, B. K., J. D. Nichols, and M. J. Conroy. 2002. Analysis and management of animal populations. Academic Press, San Diego, California, USA.

\bibitem[\citeproctext]{ref-williams2016}
Williams, P. J., and M. B. Hooten. 2016. Combining statistical inference and decisions in ecology. Ecological Applications 26:1930--1942.

\end{CSLReferences}
\end{document}